\input harvmac
\input epsf

\overfullrule=0pt
\abovedisplayskip=12pt plus 3pt minus 3pt
\belowdisplayskip=12pt plus 3pt minus 3pt
%
\def\tilde{\widetilde}

\def\to{\rightarrow}
\def\tphi{{\tilde\phi}}

\def\bigone{\hbox{1\kern -.23em {\rm l}}}
\def\ZZ{\hbox{\zfont Z\kern-.4emZ}}
\def\half{{\litfont {1 \over 2}}}
\def\cO{{\cal O}}
\def\tcO{{\tilde {\cal O}}}
\def\hcO{{\hat {\cal O}}}
\def\tcOp{{\tilde {\cal O} p}}
\def\hcOp{{\hat {\cal O} p}}
\def\cOM{{\cal OM}}
\def\tr{{\rm tr}\,}
\def\hA{{\hat A}}
\def\hcL{{\hat {\cal L}}}
\font\litfont=cmr6


\lref\bsv{M. Bershadsky, V. Sadov and C. Vafa, {\it ``D-Branes and
Topological Field Theories''}, Nucl. Phys. {\bf B463} (1996) 420,
hep-th/9511222.}
\lref\ghm{M.B. Green, J. Harvey and G. Moore,{\it ``I-brane Inflow and 
Anomalous Couplings on D-branes''}, Class. Quant. Grav. {\bf 14}
(1997) 47, hep-th/9605033.}
\lref\cheungyin{Y.E. Cheung and Z. Yin, {\it ``Anomalies, Branes and
Currents''}, Nucl. Phys. {\bf B517} (1998) 69, hep-th/9710206.}
\lref\hori{K. Hori, {\it ``Consistency Condition for Five-Brane in 
M-theory on $R^5/Z_2$ Orbifold''}, Nucl. Phys. {\bf B539} (1999) 35,
hep-th/9805141.}
\lref\witvect{E. Witten, {\it ``Toroidal Compactification Without
Vector Structure''}, JHEP {\bf 02} (1998) 006, hep-th/9712028.}
\lref\djm{K. Dasgupta, D. Jatkar and S. Mukhi, {\it ``Gravitational
Couplings and $Z_2$ Orientifolds''}, Nucl. Phys. {\bf B523} (1998)
465, hep-th/9707224.}
\lref\crapsone{B. Craps and F. Roose, {\it ``Anomalous D-brane and 
Orientifold Couplings from the Boundary State''}, Phys. Lett. {\bf
B445} (1998) 150, hep-th/9808074.}
\lref\mssone{J. Morales, C. Scrucca and M. Serone, {\it ``Anomalous 
Couplings for D-branes and O-planes''},  Nucl. Phys. {\bf B552} (1999) 
291, hep-th/9812071.}
\lref\stefan{B. Stefanski, {\it ``Gravitational Couplings of D-branes 
and O-planes''}, Nucl. Phys. {\bf B548} (1999) 275,
hep-th/9812088.}
\lref\crapstwo{B. Craps and F. Roose, {\it ``(Non)-Anomalous
D-brane and O-plane Couplings: The Normal Bundle''}, Phys. Lett. {\bf
B450} (1999) 358, hep-th/9812149.}
\lref\sstwo{C. Scrucca and M. Serone, {\it ``Anomalies and Inflow on
D-branes and O-planes''}, Nucl. Phys. {\bf B556} (1999) 197, 
hep-th/9903145.}
\lref\hanzaf{A. Hanany and A. Zaffaroni, {\it ``Issues on
Orientifolds: On the Brane Construction of Gauge Theories with $SO(2N)$
Global Symmetry''}, hep-th/9903242.}
\lref\landlop{K. Landsteiner and E. Lopez, {\it ``New Curves from
Branes''}, Nucl. Phys. {\bf B516} (1998) 273, hep-th/9708118.} 
\lref\dmlow{K. Dasgupta and S. Mukhi, {\it ``A Note on Low-Dimensional
String Compactification''}, Phys. Lett. {\bf B398} (1997) 285,
hep-th/9612188.}
\lref\senstr{A. Sen, {\it ``Strong Coupling Dynamics of Branes from
M-theory''}, JHEP {\bf 10} (1997) 002, hep-th/9708002.}
\lref\smsltz{S. Mukhi, {\it ``Dualities and the $SL(2,Z)$ Anomaly''},
JHEP {\bf 12} (1998) 006, hep-th/9810213.}
\lref\dabpark{A. Dabholkar and J. Park, {\it ``Strings on
Orientifolds''}, Nucl. Phys. {\bf B477} (1996) 701, hep-th/9604178.}
\lref\gabgreen{M. Gaberdiel and M.B. Green, {\it ``An $SL(2,Z)$ Anomaly 
in IIB Supergravity and its F-theory Interpretation''}, JHEP {\bf 11}
(1998) 026, hep-th/9810153.}
\lref\chl{S. Chaudhuri, D. Hockney and J. Lykken {\it ``Maximally
Supersymmetric String Theories in $D < 10$''}, Phys. Rev. Lett. {\bf
75} (1995) 2264, hep-th/9505054\semi 
S. Chaudhuri and J. Polchinski, {\it ``Moduli Space of CHL Strings''},
Phys. Rev. {\bf D52} (1995) 7168, hep-th/9506048\semi
W. Lerche, C. Schweigert, R. Minasian and S. Theisen, {\it ``A Note on
the Geometry of CHL Heterotic Strings''}, Phys. Lett. {\bf B424}
(1998) 53, hep-th/9711104.}
\lref\senschw{J.H. Schwarz and A. Sen, {\it ``Type IIA Dual of the
Six-Dimensional CHL Compactification''}, Phys. Lett. {\bf B357} (1995) 
323, hep-th/9507027.}
\lref\horwit{P. Horava and E. Witten, {\it ``Heterotic and type I 
string dynamics from eleven-dimensions''}, Nucl. Phys. 
{\bf B460} (1996), 506, hep-th/9510209,}
\lref\dmfive{K. Dasgupta and S. Mukhi, {\it ``Orbifolds of M-theory''},
Nucl. Phys. {\bf B465} (1996) 399, hep-th/9512196.}
\lref\witfive{E. Witten, {\it ``Five branes and M-theory on an
Orbifold''}, Nucl. Phys. {\bf B463} (1996) 383, hep-th/9512219.}
\lref\witbaryon{E. Witten, {\it ``Baryons and Branes in 
Anti-deSitter Space''}, JHEP {\bf 07} (1998) 006, hep-th/9805112.}
\lref\ejs{N. Evans, C.V. Johnson and A.D. Shapere, {\it
``Orientifolds, Branes and Duality of 4d Gauge Theories''},
Nucl. Phys. {\bf B505} (1997), 251, hep-th/9703210.} 
\lref\strom{A. Strominger, {\it ``Open $p$-branes''}, Phys. Lett. {\bf 
B383} (1996) 44, hep-th/9512059.}

{\nopagenumbers
\Title{\vtop{\hbox{hep-th/9907215}
\hbox{TIFR/TH/99-33}}}
{\vtop{\centerline{Gravitational Couplings,}
\centerline{}
\centerline{Orientifolds and M-Planes}}}
\centerline{Sunil Mukhi\foot{E-mail: mukhi@tifr.res.in}}
\vskip 3pt
\centerline{and}
\vskip 3pt
\centerline{Nemani V. Suryanarayana\foot{E-mail: nemani@tifr.res.in}}
\vskip 3pt
\centerline{\it Tata Institute of Fundamental Research,}
\centerline{\it Homi Bhabha Rd, Mumbai 400 005, India}
\vskip 5pt

\ \smallskip
\centerline{ABSTRACT}

We examine string-theory orientifold planes of various types including
the $Sp$ and $SO$-odd planes, and deduce the gravitational
Chern-Simons couplings on their world-volumes. Consistency checks are
carried out in different spacetime dimensions using various dualities,
including those relating string theory with F-theory and M-theory. It
is shown that when an orientifold 3-plane crosses a 5-brane, the jump
in the charge is accompanied by a corresponding change in the
gravitational couplings.

\Date{July 1999}
\vfill\eject}
\ftno=0

\newsec{Introduction}

Topological or Chern-Simons couplings on Dirichlet branes have played
an important role in string theory and the understanding of
dualities. These couplings involve two kinds of fields: fields arising
in the bulk spacetime theory (from the RR or NS-NS sectors) and fields
coming from the brane worldvolume theory. 

In particular, there are gravitational couplings of this kind 
involving only bulk fields, namely the spacetime metric and various 
$m$-form potentials. Schematically, they are of three types:
\eqn\couplings{
\int C\wedge \tr(R\wedge R),\qquad 
\int C\wedge \left( \tr(R\wedge R)\right)^2,\qquad 
\int C\wedge \tr(R\wedge R \wedge R \wedge R) }
In each expression, $C$ is some $m$-form gauge potential where $m$ is
such that the term has the correct rank to be integrated over
the relevant brane world-volume. The second and third terms involve
8-forms in the curvature, hence they arise only on $p$-branes for
$p\ge 7$. We will refer to these as ``8-form terms''. The first term
occurs on all D $p$-branes for $3\le p \le 9$, and will be referred to
as a ``4-form'' term. The presence of such terms was first predicted
(in a special case) in Ref.\refs\bsv\ using duality. The general form
was worked out using an anomaly inflow argument in Ref.\refs\ghm, and
extended to the case of a twisted normal bundle in Ref.\refs\cheungyin.

The above terms do not involve world-volume fields of the D-brane, so
they can be thought of as interactions among bulk fields that are
localized on the brane. From this perspective, there is no reason why
orientifold planes cannot have similar interactions residing on them.
Indeed, it was first observed in Ref.\refs\djm\ that orientifold
planes indeed have gravitational 4-form and 8-form world-volume
couplings, whose coefficients were determined using heterotic--type-I
duality. The presence of such interactions was tested in
Ref.\refs{\djm,\senstr} using dualities involving M and F theory. 

In recent times, the presence of gravitational couplings on both
branes and planes has been confirmed by explicit perturbative string
computations. In particular, as far as $\cO$-plane couplings are
concerned, in Ref.\refs\crapsone\ the gravitational 4-form couplings
predicted in Ref.\djm\ were confirmed by computing appropriate
scattering amplitudes. Refs.\refs{\mssone,\stefan}\ performed more
general computations, thereby determining the 8-form couplings as
well, and correcting a numerical error in Ref.\refs\djm. They also
noted that the $\cO$-plane gravitational couplings are summarized in
the Hirzebruch polynomial, just as the D-brane gravitational couplings
are summarized in the A-roof genus. The complete result including the
extension to the case of a nontrivial normal bundle is to be found in
Ref.\refs\mssone, while related further work was carried out in
Refs.\refs{\crapstwo,\sstwo}.

The orientifold planes considered in the above works are the ones
that give rise to $D_n$ gauge theories when branes coincide with
them. They always carry a negative amount of D-brane charge and are
usually called $\cO p^-$ planes\foot{Unfortunately, they are also
sometimes called $\cO^+$ planes\refs\witvect. We will not use this
notation.}. The purpose of this note is to examine other kinds of
orientifold planes from the perspective of world-volume gravitational
couplings. In particular, we will focus on the kind which give rise to 
$C_n$ or $Sp$-type gauge theories when branes coincide with them.
These always carry a positive charge and will be denoted $\cO p^+$ in
what follows. 

At least two other types of orientifold planes exist, but they have
been observed only in a few specific (low) dimensions. One of them,
which produces $SO(2n+1)$ type gauge groups, consists of an $\cO p^-$
plane with a single D-brane stuck on it, and will be denoted
$\tcOp$. The other type of plane is somewhat mysterious and gives rise
to $Sp$-type gauge groups, but differs from the $\cO p^+$ plane
mentioned above. Following Ref.\refs\hanzaf, we will denote it
$\hcOp$.

The D $p$-brane charge\foot{These charges are in units where a mirror
pair of D-branes has charge $+1$.} of an $\cO p^-$ plane is
$-2^{p-5}$, while for the $\cO p^+$ plane it is $2^{p-5}$. For the
$\tcOp$ plane the charge is $-2^{p-5} + \half$ in the dimensions where
it exists, while for the $\hcOp$ plane the charge is known only in a
few specific dimensions, as we will discuss below. The $\tcOp$ do not
have simple transformation properties under T-duality, while the $\cO
p^-$ and $\cO p^+$ planes simply double each time we compactify on a
circle and T-dualize. We will see in one case that assuming this also
for the $\hcO$-plane leads to a consistent result.

\newsec{Gravitational couplings on $\cO^-$ planes}

The general formulae obtained in Ref.\refs\djm\ can be summarised as
follows. An orientifold $p$-plane of type $\cO p^-$, for $3\le p
\le9$, has a gravitational 4-form coupling on its world-volume given
by 
\eqn\ominusfour{
\cO p^-:\qquad 
-{2^{p-9}\over 3} \int\, C^{RR}_{p-3}\wedge {\tr R\wedge R\over
16\pi^2} = -{2^{p-9}\over 6}\,\int\,  C^{RR}_{p-3} \wedge p_1}
where $p_1$, the first Pontryagin class of the manifold, is given by
\eqn\pone{
p_1(R) = {1\over 8\pi^2} \tr R\wedge R }
and $C^{RR}_{p-3}$ is the $(p-3)$-form Ramond-Ramond gauge potential.
For comparison, a single D$p$-brane carries the coupling 
\eqn\singledfour{
Dp: \qquad - {1\over 48} \int\,  C^{RR}_{p-3}\wedge p_1}
This formula is unambiguous in the absence of orientifold
planes. However, when an orientifold plane is present, D-branes can
only move around in mirror pairs. In this case, a single D-brane of
a mirror pair carries the coupling
\eqn\singleofpair{
-{1\over 96}\int\,   C^{RR}_{p-3}\wedge p_1}
In particular, in situations where ``half'' a D-brane is considered to
be stuck at the orientifold, this object carries the term in
Eq.\singleofpair.

For $7\le p\le 9$ there are also gravitational 8-form 
couplings\refs\djm. An $\cO p^-$ plane carries the coupling\foot{The
correct answer is to be found in the latest hep-th version of
Ref.\refs\djm, superseding the published version.}
\eqn\ominuseight{
\cO p^-:\qquad 
2^{p-9}\int\, C^{RR}_{p-7}\wedge \left({1\over 640} (p_1)^2 - {7\over
1440} p_2 \right) }
where 
\eqn\ptwo{
p_2 = {1\over (2\pi)^4}\left( -{1\over 4}\tr R^4 + {1\over 8}\,
(\tr R^2)^2 \right) }

For comparison, a single independent D $p$-brane carries the 8-form
coupling: 
\eqn\singledeight{
Dp: \qquad 
{1\over 320}\int\, C^{RR}_{p-7}\wedge \left({1\over 8}\, (p_1)^2 - {1\over
9}\, p_2 \right)}

The formulae above for gravitational couplings on D-branes and
$\cO^-$-planes can be summarised as follows\refs{\ghm,\mssone,\stefan}:
\eqn\genform{
\eqalign{
Dp: \qquad &\int \sum_i C^{RR}_i \wedge \sqrt{\hA(R)} \cr
\cO p^-: \qquad & -2^{p-5}\int \sum_i C^{RR}_i \wedge \sqrt{{\hcL}(R/4)}
\cr}}
where $\sum_i C^{RR}_i$ is the formal sum of all RR gauge potentials
in the corresponding string theory (type IIA or IIB), and $\hA$ and
$\hcL$ are the A-roof genus and Hirzebruch polynomials respectively,
given by:
\eqn\aroof{
\eqalign{
\hA(R) &\equiv 1 - {1\over 24} p_1 + {7\over 5760} (p_1)^2 - {1\over
1440} p_2 + \ldots\cr
\hcL(R/4) &\equiv 1 + {1\over 48} p_1 - {1\over 11520} (p_1)^2 + {7\over
11520} p_2 + \ldots\cr}}
Only the forms of the correct rank to be integrated over the
brane or plane worldvolume are retained in Eq.\genform. Moreover, this
formula is valid only when the normal bundle to the world-volume is
trivial. Otherwise, we must make the replacements:
\eqn\normrepl{
\eqalign{
\hA(R) &\to {\hA(R_T)\over \hA(R_N)}\cr
\hcL(R/4) &\to {\hcL(R_T/4)\over \hcL(R_N/4)}\cr}}
where $R_T$ and $R_N$ are the curvature 2-forms of the tangent and
normal bundles respectively.

\newsec{Gravitational couplings on other $\cO$ planes}

Let us now consider the other kinds of orientifold planes. Recall that
the above results were originally derived in Ref.\refs\djm\ by
appealing to heterotic/type-I duality and the low-energy effective
action of the heterotic string. The anomalous Bianchi identity
satisfied by the heterotic string 3-form $H=dB$ can be dualized to a
coupling
\eqn\dualcoup{
-\half\, \int\, {}B_6\wedge p_1(R)}
in the effective action of the heterotic string, where $B_6$ is the
dual potential to $B$, in other words $dB_6 = {}^* dB$ where $^*$ is
the Hodge dual. Dualizing to type I, $B_6$ becomes an RR 6-form
potential. Requiring that this term comes from one $\cO 9^-$ plane and
16 D9-branes, and using the results of Ref.\refs\ghm\ for
gravitational couplings on D-branes, one finds the result for the
coupling on an $\cO 9^-$ plane. Finally, the fact that $\cO^-$ planes
split into two each time we T-dualize gives the result for all $\cO^-$
planes.

The same procedure works for 8-form couplings, except that this time
the relevant terms in the effective action come not from the anomalous 
Bianchi identity but from the Green-Schwarz cancellation mechanism.

If we want to deduce the gravitational couplings on $Sp$-type
orientifold planes $\cO p^+$, we need to produce a string vacuum
containing these planes, and where the relevant terms in the effective
action are known. A convenient example is the 8-dimensional
compactification discussed in Ref.\refs\witvect, dual to a toroidal
compactification without vector structure. This has three $\cO 7^-$
planes and one $\cO 7^+$ plane, and 8 pairs of D7-branes.  It is easy
to see that the terms that were used in Ref.\refs\djm\ to find the
gravitational couplings on $\cO^-$ planes are present in this
8-dimensional model with the same coefficients as before. These terms
arise as higher derivative (or $\alpha'$) corrections in the
tree-level effective action. The 8-dimensional model in question can
be dualized to a CHL\refs\chl\ type background for the heterotic
string. This corresponds to a $Z_2$ orbifold without a twisted sector,
hence the tree-level couplings of the B-field and gravitons (which are
all invariant under $Z_2$) are unchanged.

With this result it is straightforward to find the 4-form and 8-form
couplings on the $\cO 7^+$ plane. In the present example, the 4-form
coupling on three $\cO 7^-$ planes, a $\cO 7^+$ plane and 8 pairs of
D7-branes must add up to the term in Eq.\dualcoup. Moreover, an
$\cO^+$ plane also splits into two upon T-dualizing, like the $\cO^-$
plane\refs\witvect. It follows that an $\cO p^+$ plane (which exists
for all $p\le 8$) carries the gravitational 4-form coupling:
\eqn\oplusfour{
\cO p^+: \qquad -{5.2^{p-8}\over 12}\,\int\,  C^{RR}_{p-3} \wedge p_1}
We will check this result with various dualities below. 

Another type of orientifold plane that exists for $p\le 8$ will be
denoted $\tcOp$. This is the plane that gives rise to gauge symmetries
of type SO(2n+1). It consists of a single D-brane (sometimes called a
``half'' D-brane) stuck on an $\cO^-$ plane. Hence its gravitational
4-form coupling is
\eqn\otildefour{
\tcOp: \qquad -\,{2^{p-5}+1\over 96}\,\int\,  C^{RR}_{p-3} \wedge p_1}

Next let us turn to 8-form couplings. Here, considerations analogous
to the ones discussed above lead to the expressions:
\eqn\opluseight{
\eqalign{\cO p^+:\qquad & 2^{p-8}\int\, C^{RR}_{p-7} \wedge
\left({9\over 1280}(p_1)^2 - {23\over 2880}p_2 \right)\cr
\tcOp:\qquad & \int\, C^{RR}_{p-7} \wedge \left(
{2^{p-6} + 1\over 5120} (p_1)^2 - {7.2^{p-7} + 1\over
5760} p_2\right)\cr}}
It is straightforward to check, using the above expressions, that the
9-dimensional vacuum containing three $\cO 7^-$ planes, one $\cO 7^+$
plane and 8 pairs of D7-branes, has the total 8-form term:
\eqn\totaleight{
\int\, \phi^{RR} \wedge \left( {1\over 128} (p_1)^2 - {1\over 96} p_2
\right)}
which is the dimensional reduction to 8 dimensions of the
gravitational part of the Green-Schwarz anomaly-cancelling term (here,
$\phi^{RR}$ is the Ramond-Ramond scalar of type IIB string theory).

The results above for the gravitational couplings on an $\cO
p^+$-plane can be summarised in the formula: 
\eqn\oplusgen{
\cO p^+: \qquad 2^{p-5}\int \sum_i C^{RR}_i \wedge \left(
2\sqrt{{\hA(R_T)\over \hA(R_N)}} - \sqrt{{\hcL(R_T/4)\over
\hcL(R_N/4)}}\right)}
Here, in addition to the tangent bundle contribution which we derived
above, we have conjectured the contribution of the normal bundle when
this is nontrivial.

\newsec{Consistency with Dualities}

The above formulae for gravitational couplings can be checked with
various dualities in different dimensions. These checks are not
strictly independent of the considerations that gave rise to
Eq.\oplusgen, yet it is useful to carry them out explicitly to
reassure oneself that a sensible picture emerges in every dimension.

The first check arises in 8 dimensions, where the vacuum with three $\cO
7^-$ planes and one $\cO 7^+$ plane was argued\refs\witvect\ to be
dual to a special F-theory compactification with a ``frozen'' $D_8$
singularity. Macroscopically, this is identical to a conventional
$D_8$ singularity which can split into 10 generic singular fibres. The
4-form gravitational couplings on a 7-brane or 7-plane are $SL(2,Z)$
invariant, hence we expect that the world-volume gravitational
couplings of the frozen $D_8$ singularity are equal to 10 times the
worldvolume couplings of a D7-brane. From Eq.\oplusfour, with $p=7$,
we find that the coefficient of the coupling is $-{5\over 24}$, which
is 10 times the coupling of a single D7-brane, $-{1\over 48}$, as
expected.

In 7 dimensions, the $\cO 6^+$ plane has similarly been related to
M-theory with a ``frozen'' $D_4$ singularity\refs\landlop. For us, the
key feature of this otherwise mysterious singularity is that it is
geometrically the same as an $\cO 6^-$ plane with four D6-branes on
top of it. In this dimension, the gravitational coupling on an $\cO
6^-$ plane and on a D6-brane was re-derived from M-theory in
Ref.\senstr. This derivation proceeds by starting with the well-known
``one-loop'' gravitational coupling $\int C_3\wedge I_8(R)$ in
eleven-dimensional M-theory, where 
$I_8(R)= {1\over 24}\left(p_2 - {1\over 4} (p_1)^2\right)$. 
Integrating this over the 4-dimensional manifold representing the
transverse space to a 6-brane or 6-plane, we get the induced
gravitational coupling on the corresponding object.

Both for a 2-centre Euclidean Taub-NUT space, which is the transverse
space to a D6-brane, and for the Atiyah-Hitchin space, transverse to
an $\cO 6^-$-plane, it is known that $\int p_1 = 2$. Thus, by the
above arguments, for the space transverse to an $\cO 6^+$-plane we
must have $\int p_1 = 10$. This gives rise to the induced
gravitational coupling 
\eqn\indosix{
\cO 6^+: \qquad -{5\over 48}\,\int\,  C_3 \wedge p_1}
which agrees with Eq.\oplusfour\ for $p=6$.

In 5 dimensions, we can use the beautiful results of Hori\refs\hori\
relating various types of $\cO 4$-planes to M-theory
orientifolds\refs{\horwit,\dmfive,\witfive}. The $\cO 4^-$ plane is
relatively straightforward, as it corresponds in M-theory to an
orientifold 5-plane (which we will refer to as an $\cOM$ 5-plane). The
$\cO 4^+$ plane is, according to Ref.\refs\hori, an $\cOM$-plane with
a pair of M-theory 5-branes stuck to it. Now, neither the M5-brane nor
the $\cOM$ 5-plane carry gravitational couplings\refs\smsltz. However,
being chiral objects, they carry an analogue of the $SL(2,Z)$ anomaly
of Ref.\refs\gabgreen. It was argued in Ref.\refs\smsltz, by analogy
with a similar mechanism discussed in Ref.\dmlow, that after
compactifying on a circle the chirality of these objects gives rise to
the appropriate gravitational couplings on the world-volumes of
D4-branes and $\cO 4^-$-planes. It follows that the gravitational
coupling on the $\cO 4^+$ plane is equal to that induced by an $\cOM$
5-plane and a pair of $M5$-branes, which in turn is equal to the sum
of gravitational couplings on an $\cO 4^-$-plane and a pair of
D4-branes. This leads to the coupling 
$-{5\over 192}\int A^{RR} \wedge p_1$, 
in agreement with Eq.\oplusfour\ for $p=4$.

The case of $\tcO$ 4-planes, which give rise to $SO(2n+1)$ gauge
groups when branes coincide with them, is rather different. According
to Ref.\refs\hori, these arise in M-theory by combining the
orientifolding operation in five directions with a half-shift along
the M-direction $x^{10}$. This action has no fixed points, hence from
the M-theory point of view there is no orientifold plane at
all. However, in the type IIA limit (as the radius of the $x^{10}$
direction goes to zero) the half-shift is not geometrically visible
and we seem to have an orientifold 4-plane. In fact the half-shift is
now realized as half a unit of Ramond-Ramond flux $\int A^{RR}$ along
any curve that passes through the orientifold plane. 

While it would be desirable to propose an M-theoretic origin for the
gravitational couplings on the $\tcO$ 4-planes (analogous to
chirality, discussed above for the $\cO 4^\pm$ planes), this does not
seem to be straightforward, precisely because in the M-theory limit
this 4-plane is not an orientifold at all. However, it is interesting
to note that this orientifold 4-plane is closely related to an
uncharged orientifold 8-plane that has been discussed in the
literature\refs\dabpark.  Consider the compactification of M-theory to
9 spacetime dimensions on a Klein bottle, defined as the quotient of a
2-torus labelled by $x^9,x^{10}$ by the action 
$x^9\to -x^9, x^{10}\to x^{10} + \pi R^{10}$. 
From the M-theory point of view this has no
fixed points, but in the string theory limit one finds two fixed
planes at $x^9 = 0,\pi R^9$, with the half-shift in $x^{10}$ being
realized as half a unit of $A^{RR}$-flux through each fixed
plane. These two planes are neutral and the vacuum is therefore
``braneless''. According to Ref.\refs\witvect, this model is dual to
another braneless orientifold, without RR flux, containing one 
$\cO 8^-$ and one $\cO 8^+$ plane. Hence the gravitational coupling 
on this exotic $\cO$ 8-plane is the average of the couplings on the 
$\cO 8^-$ and $\cO 8^+$ planes, or:
\eqn\neutraleight{
{\rm neutral}~\cO 8:\qquad -{1\over 4}\int\, C^{RR}_5 \wedge p_1}
While the neutral 8-planes carry RR flux and are therefore difficult
to study in perturbation theory (like the orbifold planes in
Ref.\refs\senschw), the dual vacuum containing an $\cO 8^-$
and an $\cO 8^+$ plane can be described very precisely as
a perturbation series\refs\witvect.

One can analogously realize $\tcO$ 4-planes in a braneless vacuum by
compactifying M-theory to 5 spacetime dimensions on a ``Klein
six-bottle'', defined as the quotient of the 6-torus
$x^5,\ldots,x^{10}$ by the action $(x^5,\ldots x^9)\to -(x^5,\ldots
x^9)$, $x^{10}\to x^{10} + \pi R^{10}$. This time, in the string
theory limit there are 32 neutral orientifold planes, each carrying
half a unit of RR flux. According to Ref.\refs\hori, this vacuum is
dual to the string theory vacuum discussed in Ref.\witfive\ where
there are 32 $\tcO 4$-planes, each consisting of ``half'' a D4-brane
stuck to an $\cO 4^-$-plane. It follows that, like the neutral $\cO
8$-planes, the neutral $\tcO 4$-planes carry gravitational couplings
\eqn\neutralfour{
\tcO 4:\qquad -{1\over 64} \int\, A^{RR} \wedge p_1 }
in agreement with Eq.\otildefour.

We can also examine the exotic $Sp$-type 4-plane (which we denoted
$\hcO 4^+$). This is supposed to arise\refs\hori\ as an M5-brane
wrapped on the cylinder over which the $\tcO 4$-plane is smeared. In
the string theory limit this brane will descend to half a D4-brane,
hence the coupling on this 4-plane is:
\eqn\exoticsp{
\hcO 4:\qquad -{5\over 192}\int\, A^{RR} \wedge p_1 }
which is the same as we found for the $\cO 4^+$-plane. Thus it appears 
that both types of $Sp$-planes have the same gravitational coupling,
as well as the same 4-brane charge.

Finally we come to 4 spacetime dimensions, or orientifold
3-planes. Here we find the pleasant result, from
Eqs.\oplusfour,\otildefour\ and \exoticsp\ (in the last case we assume
that the $\hcO 4$-plane also doubles on compactification and
T-duality) that three different types of orientifold 3-planes have the
same gravitational coupling:
\eqn\threetypes{
\cO 3^+, \tcO 3, \hcO 3:\qquad -{5\over 384}\int\, \tphi\wedge p_1}
consistent with the fact\refs\witbaryon\ that these three types of
$\cO 3$-planes are permuted into each other by $SL(2,Z)$ S-duality.

\newsec{Intersecting Orientifolds and Branes}

An interesting application of the above considerations can be found in
a system where an orientifold plane intersects a brane. Here we will
consider the case when an orientifold 3-plane intersects an NS
5-brane in type IIB string theory. The worldvolume directions are
$(x^1,x^2,x^3)$ for the 3-plane and $(x^1,x^2,x^4,x^5,x^6)$ for the
5-brane, so they intersect over a 2-brane in the $(x^1,x^2)$
directions. This is a supersymmetric configuration.
\bigskip

\centerline{\epsfbox{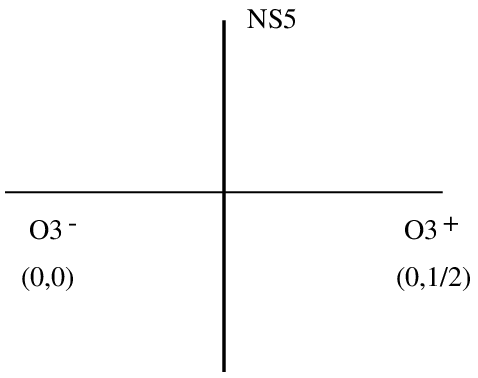}}\nobreak
\centerline{Fig. (1): $\cO 3$-plane with $\theta_{RR}=0$ intersecting 
an NS5-brane.}
\bigskip

As shown in Ref.\witbaryon, an $\cO 3^-$ plane intersecting an
NS5-brane turns into an $\cO 3^+$ plane on the other side\foot{This
phenomenon was first discovered in Ref.\refs{\ejs} for $\cO
4$-planes.}. The latter is the standard $Sp$-type plane that we have
discussed in previous sections. The particular property of $\cO
3$-planes is that the space transverse to them contains a 2-cycle (an
$RP^2$) on which the fluxes $\theta_{NS} =\int B_{NS}$ and
$\theta_{RR} = \int B_{RR}$ can take two discrete values, 0 and
$\half$. The effect of crossing an NS5-brane is that
\eqn\crossns{
(\theta_{RR},\theta_{NS}) \to (\theta_{RR},\theta_{NS}+ \half)}
where the $\theta$-values are understood to be mod 1. Hence in Fig.1,
the plane on the left is of type $(0,0)$ and the one on the right is
$(0,\half)$. 

We will now see that the world-volume gravitational couplings on the
orientifold planes jump across the NS5-brane. Consider an $RP^5$
enclosing the $\cO 3^-$ on the left. It contains the $RP^2$ cycle on
which we evaluate $(\theta_{RR}, \theta_{NS})$ to be $(0,0)$. Now move
this $RP^5$ towards the right. It intersects the NS5-brane in an
$RP^2$, enclosing the 2-brane intersection region denoted in the
figure by a point. Recall that the 3-brane charge of the orientifold
planes jumps across the NS5-brane, from $-{1\over 4}$ to $+{1\over
4}$. It is amusing to note that this is very much like branes ending
on branes\refs\strom, in that a net 3-brane charge is deposited on the 
5-brane. This is usually interpreted as the fact that the intersection 
region acts like a magnetic source in the 5-brane world-volume,
carrying the net charge deposited on the 5-brane.

In the present case, this means that the (twisted) world-volume 
$U(1)$ gauge field strength on the NS5-brane must be excited, with
\eqn\worldf{
\int_{RP^2}\, F = \half}
(in the covering space, the charges on the planes would be doubled and 
we would have $\int F=1$). Now let us consider the world-volume
couplings on the NS5-brane:
\eqn\nsworld{
\int\, F \wedge D^+ - {1\over 48}\int\, \tphi\,F \wedge p_1(R) }
In the presence of the world-volume flux through $RP^2$, these give
rise to co-dimension 2 couplings:
\eqn\codimtwo{
\half \int\, D^+ - {1\over 96}\int\, \tphi\, p_1(R) }
From the geometry of the problem, it follows that the world-volume
couplings on the $\cO 3^-$ plane on the left, added to the above
terms, must equal the world-volume couplings of the $\cO 3^+$ plane on 
the right. For the $\cO 3^-$ plane, we have the couplings
\eqn\othreemin{
-{1\over 4} \int\, D^+ - {1\over 384}\int\, \tphi\, p_1(R) }
Adding this to Eq.\codimtwo, we find the result
\eqn\othreepl{
{1\over 4} \int\, D^+ - {5\over 384}\int\, \tphi\, p_1(R) }
This exhibits the fact that the $\cO 3^+$ plane has 3-brane charge
$+{1\over 4}$ as expected, and also a gravitational 4-form coupling
with coefficient $-{5\over 384}$ as predicted in previous sections. 

Note that this jump in the world-volume coupling is not a simple
consequence of crossing the NS5-brane, as it only takes place when a
world-volume flux is excited. For a different example, consider the
$\cO 3$-plane of type $(\half,0)$, which is the $SO$-odd orientifold
plane. When this intersects an NS5-brane, the plane on the right is
the $\cO 3$-plane of type $(\half,\half)$, which is the ``exotic'' $Sp$
plane.
\bigskip

\centerline{\epsfbox{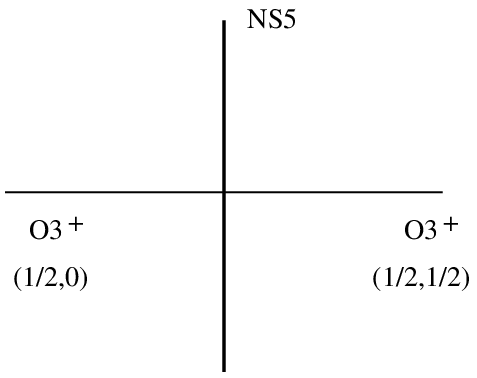}}\nobreak
\centerline{Fig. (2): $\cO 3$-plane with $\theta_{RR}=\half$ 
intersecting an NS5-brane.}
\bigskip

In this case, the orientifold planes on the left and right of the
figure both have 3-brane charge $+{1\over 4}$. Hence, even though the
value of $\theta_{NS}$ jumps as before, there is no jump in the
3-brane charge. It follows that no net charge is deposited on the
NS5-brane, hence no world-volume flux is excited. This in turn tells
us that the gravitational coupling does not jump in this case, in
accordance with the results of the previous section.

\bigskip\medskip

\noindent{\bf Acknowledgements:} 

We are grateful to Ashoke Sen for a useful correspondence.

\bigskip

\listrefs    
\end